\documentclass[%
reprint,
 amsmath,amssymb,
 aps,
]{revtex4-2}

\usepackage{graphicx}
\usepackage{dcolumn}
\usepackage{bm}
\usepackage{multirow}
\usepackage{caption}
\usepackage{blindtext}
\usepackage{appendix}
\usepackage{graphicx}
\usepackage{amsmath}
\usepackage{xcolor}
\usepackage{ulem}

\begin{document}
\title{Significance of high charge state of projectile ions inside the target and its role on electron capture leading to target ionization phenomenon }
\author{Soumya Chatterjee$^1$, Prashant Sharma$^2$, C.C. Montanari$^3$, D. Mitra$^1$ and T. Nandi$^{4*}$}
\affiliation{$^1$Department of Physics, University of Kalyani, Kalyani, Nadia-741235, WB, India. }
\affiliation{$^2$Department of Particle and Astrophysics, Weizmann Institute of Science, Rehovot 76100, Israel}
\affiliation{$^3$ Instituto de Astronomía y Física del Espacio, CONICET and Universidad de Buenos Aires, Buenos Aires, Argentina}
\affiliation{$^{4}$Inter-University Accelerator Centre, Aruna Asaf Ali Marg, Near Vasant Kunj, New Delhi-110067, India.}
\thanks {Email:\hspace{0.0cm} nanditapan@gmail.com. Present address: 1003 Regal, Mapsko Royal Ville, Sector-82, Gurgaon-122004, India.}

\begin{abstract}
The K x-ray spectra of different targets (Cu, Zn, and Ge) induced by 3 to 5 MeV/u Si projectile ions have been measured to determine the K-shell ionization cross-section. A significant difference is observed between the measurements and theoretical estimates,   { with the latter being about 50\% below the experimental results. This underestimation} is attributed to the charge-exchange from target K-shell to projectile K- and L-shells.  Such observation can only be possible if the projectile ions attain up to H- and He-like charge states. Corresponding projectile charge state fractions have been evaluated from the Lorentzian charge state distribution, where mean charge state is taken from the Fermi gas model [Phys. Rev. Lett. 30, 358 (1973)] and width from the Novikov and Teplova approach [Phys. Lett. A378, 1286–1289 (2014)]. The sum of the direct ionization cross-section and K-K + K-L capture cross-sections gives a good agreement with the measured cross-sections. Furthermore, we have validated this methodology with available data for Si-ion on Ti target. Such results may be useful in many solid target based applications. \\

\end{abstract}
\maketitle
\section{Introduction}
\indent The study of ionization dynamics of target atoms by energetic heavy ions is critical in several fields of research such as material analysis, material engineering, atomic and nuclear physics, accelerator physics,  {biophysics}, medical science, etc. The precise data of ionization cross-section of target atoms are required in case of heavy-ion application in  {particle-induced} x-ray emission (PIXE) \cite{miranda2007x} and in heavy-ion  {tumor} therapy \cite{kraft2000tumor}. Appropriate knowledge of K-shell ionization  {is essential} to determine the elemental concentration during PIXE analysis and to estimate the direct damage of the  {tumor} by the projectiles. Besides the target ionization, the effect of secondary electrons during heavy ion impact in the patient's body is very significant  {leading} to much greater damage than the direct damage by the incident ions. The secondary electron yield is found to be proportional to the rate of energy loss of the incident particles \cite{sternglass1957theory}, which again depends on the projectile charge state inside the target \cite{vager1976polarization,lifschitz2004effective}. Further, knowledge of the charge state of the projectile ions inside the target imparts a crucial role in electron capture processes  leading to the inner shell ionization in the target atoms \cite{lapicki1980electron}. \\
\indent Though a monoenergetic beam with a fixed charge state is passed through the target material, a charge state distribution (CSD) of the projectile ions is manifested inside the target and being altered at the exit surface of the target before we measure it with the   {electromagnetic technique. This technique employs a dipole magnet, kept away from the target chamber, for dispersing the different charge states and a position sensitive detector at the focal point to catch up all the dispersed ions}. CSD of the coronal mass ejections has been measured by observatory borne charge analysers \cite{stone1998advanced,gloeckler1998investigation} and theoretically studied \cite{lepri2010direct}. Highly charged ions (HCIs) are prevalent in the inertial confinement fusion (ICF) \cite{lindl1995development}. The recent development of free electron laser has led to producing HCIs during x-ray-atom interactions and the CSD has been measured and theoretically interpreted \cite{young2010femtosecond,vinko2012creation,rudek2012ultra}. The subject of this charge changing processes is thus  {significant}.  
\begin{figure*}
\centering
\includegraphics[scale=0.33]{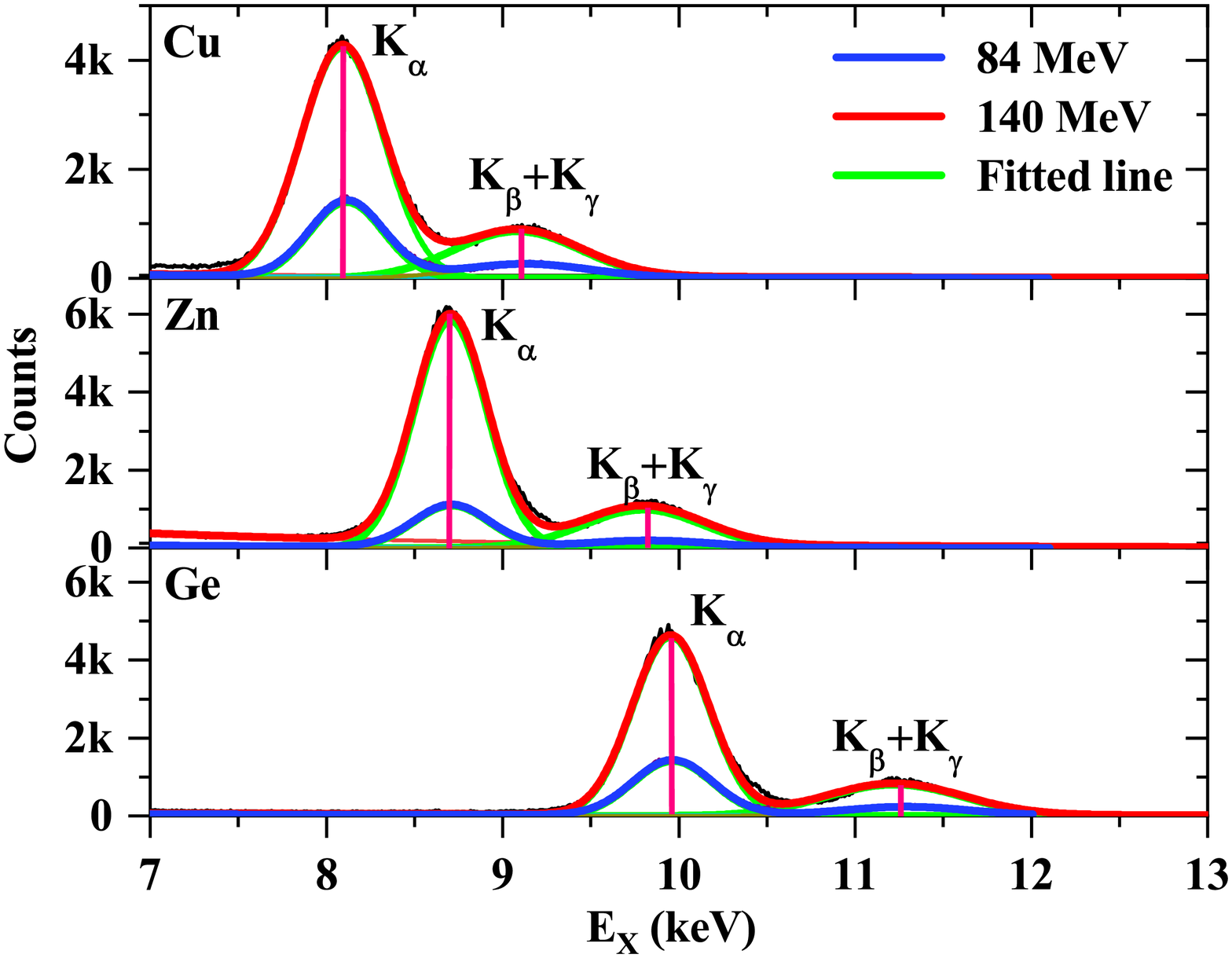}\includegraphics[scale=0.33]{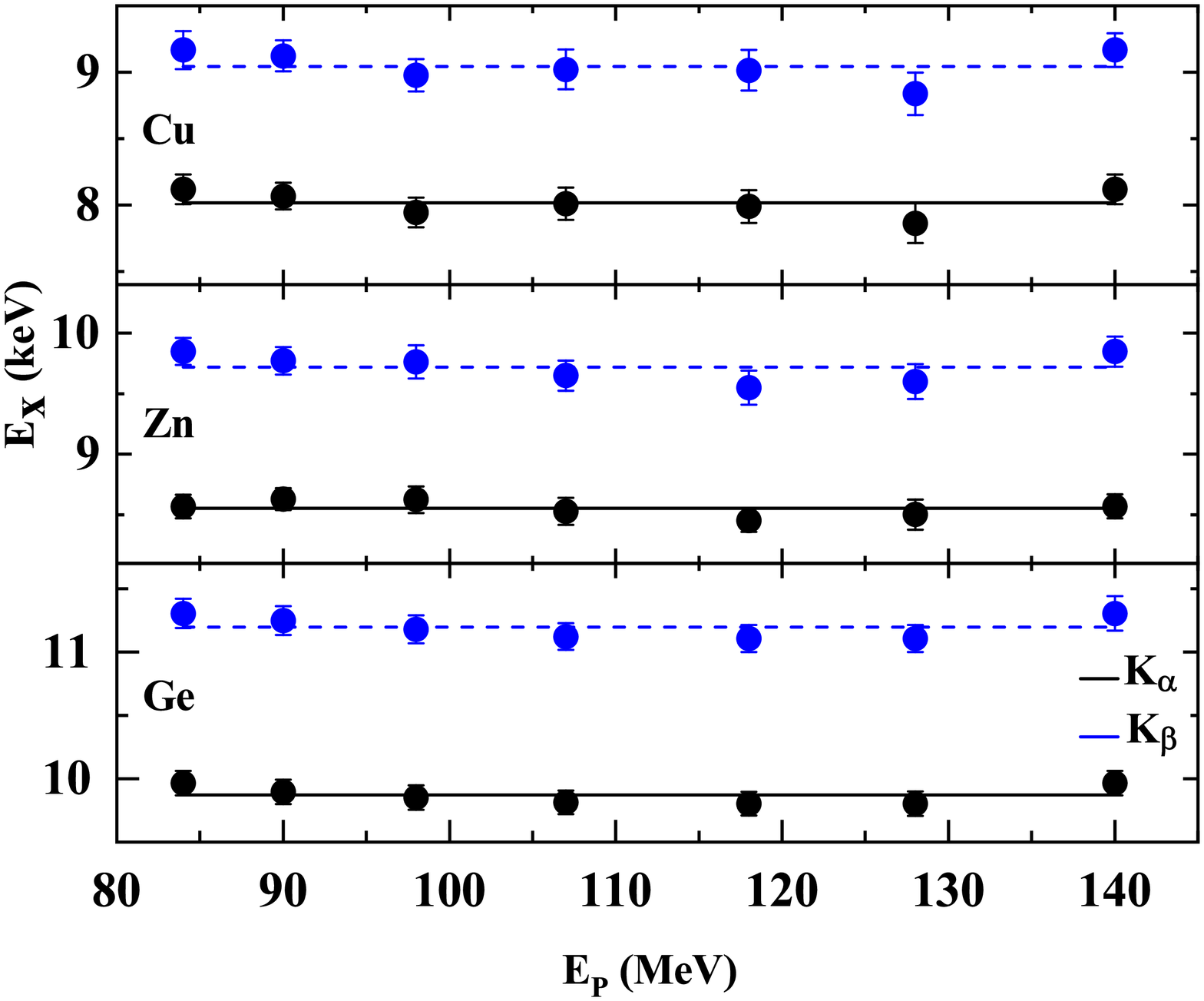}
\caption{Typical K x-ray spectra of natural Cu, Zn, and Ge targets when bombarded with 84 and 140 MeV $^{28}$Si ions (left panel) and the x-ray peak energy shift in these targets as a function of the ion-beam energies of $^{28}$Si ions (right panel). The solid and dotted orange lines in right panel show the mean value of the $K_\alpha$ and $K_\beta$ peak energies.}
\label{Spectra}
\end{figure*}

Inner shell ionization by ion impact has been investigated in the laboratory with the availability of the accelerators since the 1950's \cite{stier1954charge}. A significant difference of ionization in gas and solid target was measured away from the target using a charge  {analyzer} \cite{moak1968h}. A vital role inside a target was put through a model associating the Auger processes which occur after the ions leave the solid \cite{betz1970charge}. Nevertheless, any direct measurement of the ionization phenomena inside the target was not possible until a couple of years ago \cite{nolen2013charge}. Recently, it has been done using the x-ray spectroscopy technique \cite{sharma2016x}. The charge state of the projectile ion ($q$) in the beam-foil plasma created due to ion-solid interactions \cite{sharma2016experimental} is considerably higher than the measured ionic state outside the target \cite{sharma2019disentangling} because of electron capture phenomena from the exit surface \cite{nandi2008formation}.
 {The problem of accurate charge states of ions inside solids is still a challenge in stopping power of ions in matter \cite{montanari2017iaea}.} Such interesting features have not been exploited in the foil stripper technology  {to} date. \\
\begin{figure*}
\centering
\includegraphics[scale = 0.62]{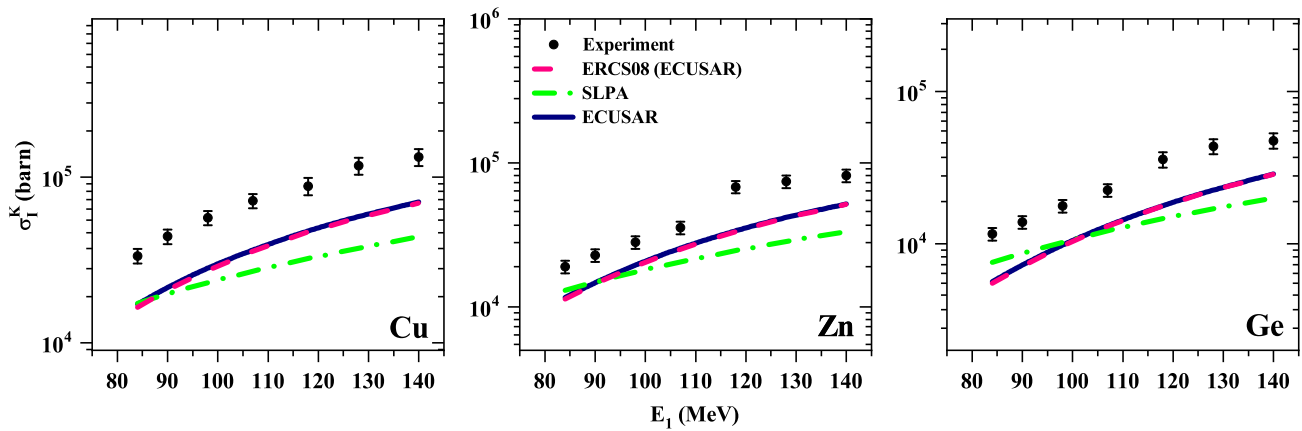}
\caption{Comparison of experimental K shell ionization cross-sections of different targets bombarded by the $^{28}{Si}$ ions as a function of ion-beam energies with the direct ionization cross-sections from ECUSAR \cite{lapicki2004effects}, SLPA \cite{montanari2013theory} and ERCS08 \cite{horvat2009ercs08}.
\label{Expt and DI theories}}
\end{figure*}
\begin{table}[]
\centering
\caption{\label{EXPERIMENTAL CROSS-SECTION} Measured K shell production cross-sections ({$\sigma_k^x$}) for Cu, Zn, and Ge targets with corresponding multiple ionization probabilities, modified fluorescence yields for multiple ionization \cite{lapicki2004effects} and ionization cross-sections ($\sigma_k^I$) by $^{28}$Si ions at different energies $(E_1)$ (MeV). Single vacancy fluorescence yields $(\omega_k^0)$ for Cu, Zn, and Ge are 0.454, 0.486, and 0.546, respectively \cite{krause1979atomic}. The cross-sections are in units of barns/atom.}
\begin{tabular}{|c|c|c|c|c|}
\hline
\multicolumn{5}{|l|}{\hspace{3cm}\textbf{Cu}}                                         \\ \hline
\textbf{$E_1$}   & \hspace{0.3cm}\textbf{$\sigma_k^x$} & \hspace{0.4cm}P     & $\omega_{k}$ & \hspace{0.3cm}$\sigma_k^I$ \\ \hline
84  & 27389$\pm$1415           & 0.817 & 0.819 & 33400$\pm$3340           \\ \hline
90  & 34211$\pm$1767           & 0.766 & 0.780 & 43800$\pm$4380           \\ \hline
98  & 41910$\pm$2160           & 0.705 & 0.738 & 56800$\pm$5680           \\ \hline
107 & 50481$\pm$2580           & 0.649 & 0.703 & 71800$\pm$7180           \\ \hline
118 & 69875$\pm$5329           & 0.590 & 0.670 & 88200$\pm$8820         \\ \hline
128 & 89307$\pm$6786           & 0.547 & 0.647 & 117000$\pm$11700         \\ \hline
140 & 97236$\pm$7343           & 0.501 & 0.625 & 132000$\pm$13200         \\ \hline
\multicolumn{5}{|l|}{\hspace{3cm}\textbf{Zn}}                                         \\ \hline
84  & 15459$\pm$825             & 0.785 & 0.815 & 19000$\pm$1900           \\ \hline
90  & 17851$\pm$947            & 0.736 & 0.782 & 22800$\pm$2280           \\ \hline
98  & 21025$\pm$1107           & 0.680 & 0.747 & 28100$\pm$2810           \\ \hline
107 & 25542$\pm$1337           & 0.626 & 0.717 & 35600$\pm$3560           \\ \hline
118 & 46908$\pm$3675           & 0.571 & 0.688 & 68200$\pm$6820           \\ \hline
128 & 49596$\pm$3865           & 0.529 & 0.668 & 74300$\pm$7430           \\ \hline
140 & 53002$\pm$4115           & 0.486 & 0.648 & 81800$\pm$8180           \\ \hline
\multicolumn{5}{|l|}{\hspace{3cm}\textbf{Ge}}                                         \\ \hline
84  & 9416$\pm$509             & 0.724 & 0.813 & 11600$\pm$1160            \\ \hline
90  & 10959$\pm$589             & 0.679 & 0.789 & 13900$\pm$1390           \\ \hline
98  & 13560$\pm$724             & 0.632 & 0.766 & 17700$\pm$1770           \\ \hline
107 & 16684$\pm$877            & 0.581 & 0.742 & 22500$\pm$2250           \\ \hline
118 & 30520$\pm$2397           & 0.534 & 0.721 & 35800$\pm$3580           \\ \hline
128 & 36339$\pm$2846           & 0.495 & 0.704 & 43700$\pm$4370           \\ \hline
140 & 38722$\pm$2994           & 0.456 & 0.688 & 47600$\pm$4760           \\ \hline
\end{tabular}
\end{table}
\indent  {Research on} inner shell ionization of target atoms  {has been} carried out for a long time with light as well as heavy projectiles  {(see for example
\cite{benka1978tables,orlic1994experimental,kadhane2003k,lapicki2005status,zhou2013k,msimanga2016k,kumar2017shell,oswal2018x,hazim2020high,oswal2020experimental,miranda2020total})}. 
  {This has enabled the research community to study processes} like ionization, excitation, multiple ionization \cite{lapicki2004effects}, radiative decay, Auger-decay \cite{dahl1976auger}, changes in atomic parameters, intrashell coupling effect \cite{pajek2003multiple,sarkadi1981possible}, etc at different energy regimes. Such processes occurring inside \cite{sharma2016x} as well as at the target surface \cite{nandi2008formation} change the initial charge state of the projectile to several charge states. Thus, a CSD is measured by any set up placed away from the target. It is worth noting that the CSD depends on the initial parameters of the projectile ion (energy, initial charge state and atomic number) as well as target characteristics (thickness, density, and atomic number). Various groups have reported the CSDs using the techniques like the electromagnetic method \cite{MAIDIKOV1982295}, recoil separator \cite{LEINO1995653,KHUYAGBAATAR201240}, TOF \cite{DICKEL2015137} and Coincident Rutherford   {Backscattering Spectrometry (CRBS) \cite{sa2011coincident}} to obtain the CSDs of the projectile outside the target, which has the combined effect of charge exchange processes in the bulk as well as the surface of the target. However, these techniques fail to separately measure the CSD of the projectile inside and outside the solid target \cite{lifschitz2004effective}. Theoretical studies include only the CSD outside the target as seen in several reviews \cite{RevModPhys.30.1137, WITTKOWER1973113,SHIMA1986357,SHIMA1992173}, which include empirical models such as Bohr model \cite{bohr1941velocity}, Betz model \cite{betz1970charge}, Nikolaev-Dmitriev model \cite{nikolaev1968equilibrium}, To-Drouin model \cite{to1976etude}, Shima-Ishihara-Mikumo model \cite{shima1982empirical}, Itoh model \cite{itoh1999equilibrium}, 
and Schiwietz-Grande model \cite{schiwietz2004femtosecond}. However,  theoretical models for estimating the CSDs inside the target   {are scarce, and this may be because of no concrete experimental guidelines up to now.}\\
\begin{figure*}
\centering
\includegraphics[scale=0.62]{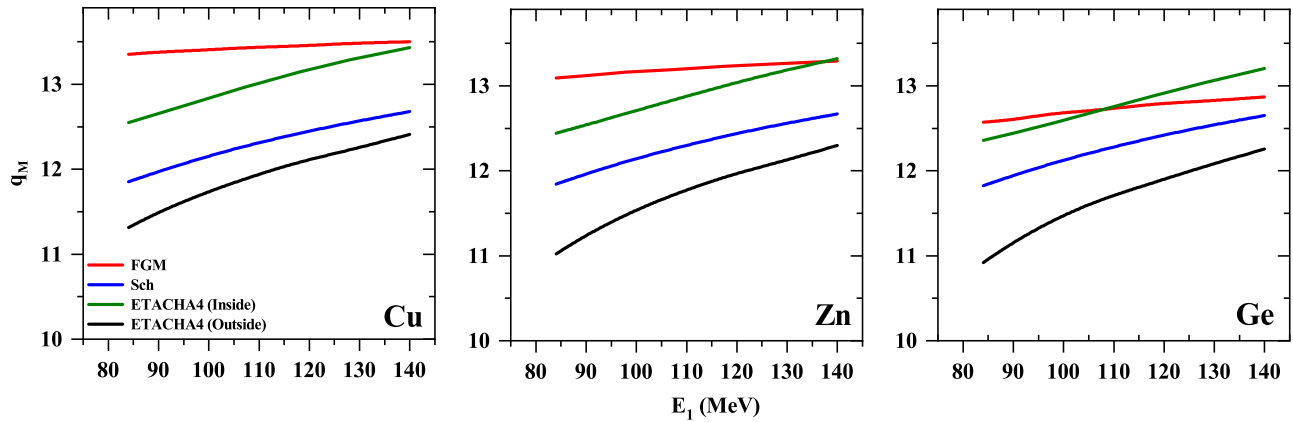}
\caption{Mean charge state of $^{28}$Si ions inside the targets Cu, Zn, and Ge as predicted by the Fermi Gas Model (FGM) \cite{brandt1973dynamic} and ETACHA4 (inside) are plotted against the incident energies. The same outside the targets as predicted by the Schiwietz-Grande model (Sch) \cite{schiwietz2001improved} and ETACHA (outside) are shown as a function of the incident energies.}
\label{Qm}
\end{figure*}
%
\begin{figure*}
\centering
\includegraphics[scale = 0.32]{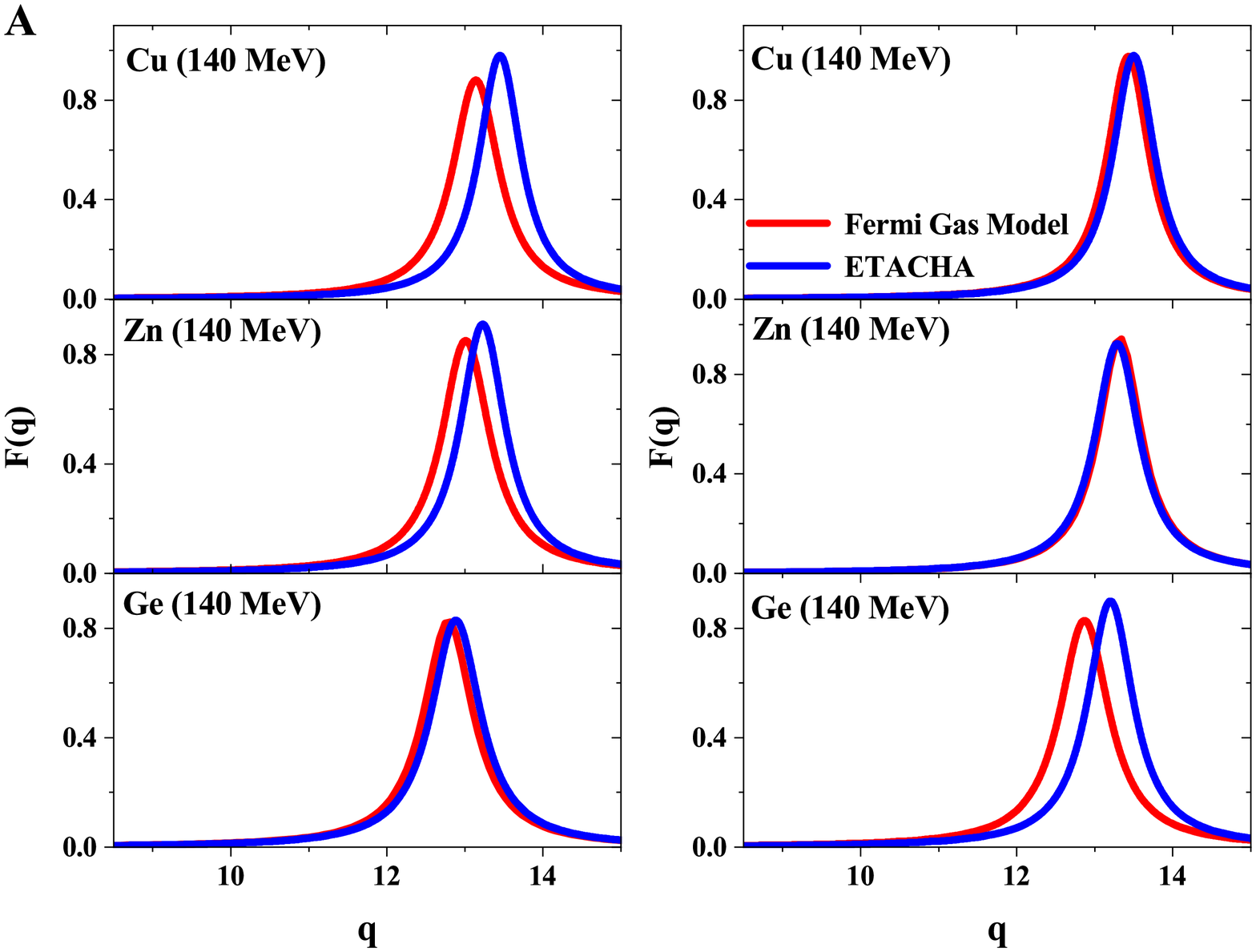}\hspace{0.5cm} \includegraphics[scale = 0.32]{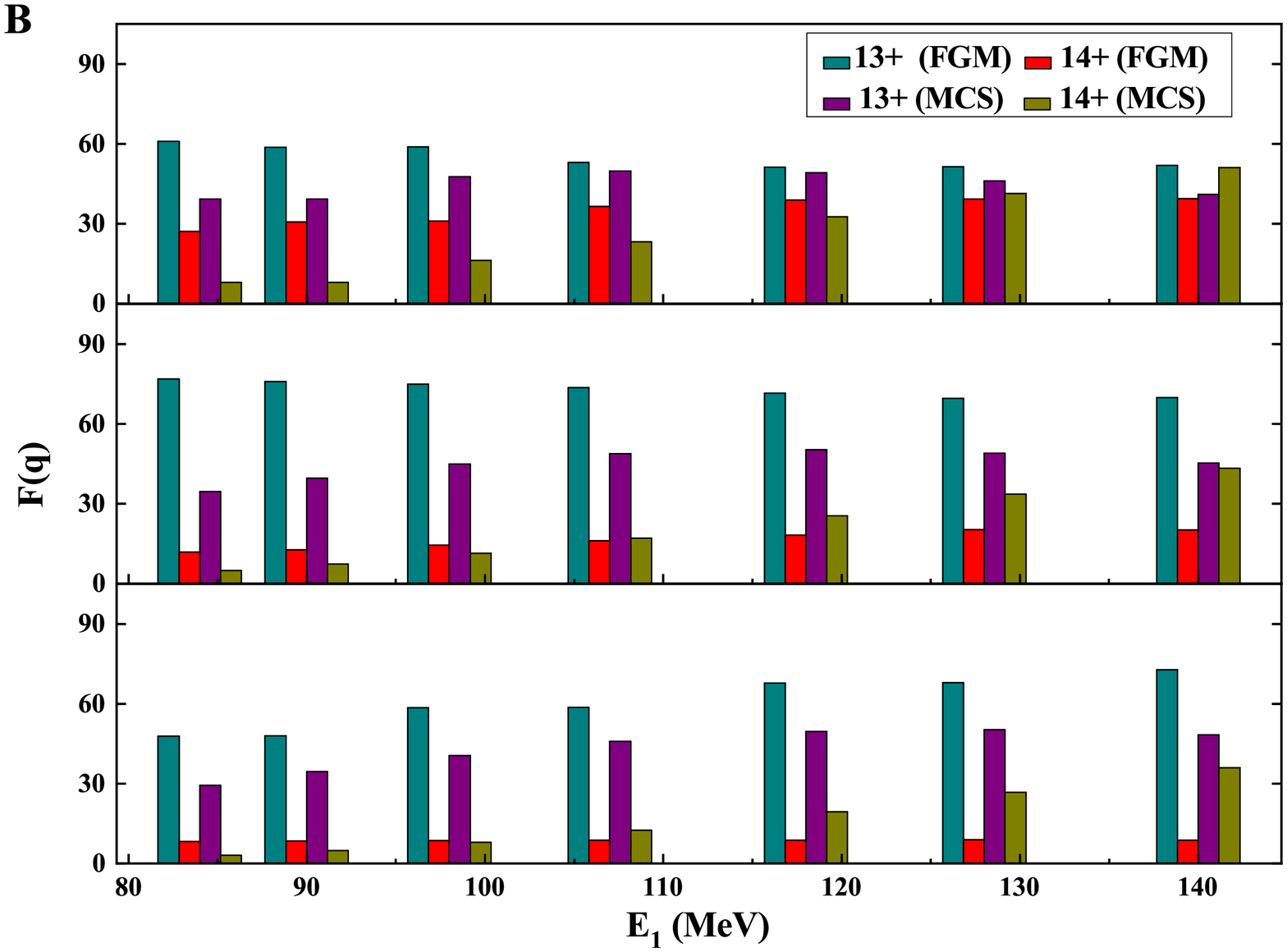}
\caption{A. Charge state distributions in side different targets according to the Fermi Gas Model (FGM) \cite{brandt1973dynamic} and ETACHA4(inside) \cite{ETACHA4} at two select energies, B. Charge state fraction (F(q)) chart for q = 13+ and 14+ inside the targets as a function of beam energies are also shown according to the Fermi Gas Model (FGM) \cite{brandt1973dynamic} and ETACHA4(inside) \cite{ETACHA4}.}
\label{CSD BAR}
\end{figure*}
\begin{table}
\centering
\caption{\label{CSD FRACTION}Charge state fraction in \% ($F(q)$) of Si$^{13+}$ and Si$^{14+}$ obtained from FGM \cite{brandt1973dynamic}, ETACHA4 \cite{ETACHA4} and ERCS08 \cite{horvat2009ercs08} in different target elements and at various kinetic energies of Si-ion beam. Note that FGM amd ETACHA4 represent $F(q)$ inside the target, while ERCS08 gives the same outside the target. }
\begin{tabular}{|c|c|c|c|c|c|c|}
\hline
\multirow{2}{*}{$E_1$} & \multicolumn{2}{l|}{\hspace{0.9cm}FGM} & \multicolumn{2}{l|}{\hspace{0.5cm}ETACHA} & \multicolumn{2}{l|}{\hspace{0.5cm}ERCS08} \\ \cline{2-7} 
                       & $F(13+)$      & $F(14+)$        & $F(13+)$         & $F(14+)$         & $F(13+)$         & $F(14+)$          \\ \hline
\multicolumn{7}{|l|}{\hspace{3.6cm}\textbf{Cu}}\\ \hline
84 & 61.0 & 27.0 & 39.2 & 7.9 & 19.5 & 2.4 \\ \hline
90 & 58.6 & 30.7 & 39.2 & 7.9 & 22.9 & 3.1 \\ \hline
98 & 58.8 & 30.9 & 47.6 & 16.3 & 27.5 & 4.3 \\ \hline
107 & 53.0 & 36.4 & 49.7 & 23.1 & 32.5 & 5.8 \\ \hline
118 & 51.2 & 38.9 & 49.1 & 32.6 & 37.9  & 7.8 \\ \hline
128 & 51.3 & 39.2 & 46.1 & 41.3 & 42.2 & 9.9 \\ \hline
140 & 51.8 & 39.4 & 41 & 51.1 & 46.4 & 12.6 \\ \hline
\multicolumn{7}{|l|}{\hspace{3.6cm}\textbf{Zn}} \\ \hline
84 & 76.8 & 11.9 & 34.5 & 4.9 & 19.2 & 2.3 \\ \hline
90 & 75.9 & 12.6 & 39.5 & 7.2 & 22.7 & 3.0 \\ \hline
98 & 74.8 & 14.4 & 44.8 & 11.3 & 27.2 & 4.2 \\ \hline
107 & 73.6 & 16.0 & 48.7 & 17.0 & 32.1 & 5.6 \\ \hline
118 & 71.4 & 18.1 & 50.3 & 25.4 & 37.6 & 7.7 \\ \hline
128 & 69.5 & 20.2 & 49.0 & 33.6 & 41.9 & 9.7 \\ \hline
140 & 69.8 & 20.1 & 45.2 & 43.3 & 46.1 & 12.3 \\ \hline
\multicolumn{7}{|l|}{\hspace{3.6cm}\textbf{Ge}} \\ \hline
84 & 47.8 & 8.3 & 29.4 & 3.2 & 18.7 & 2.2 \\ \hline
90 & 47.9 & 8.4 & 34.5 & 4.8 & 22.1 & 2.9  \\ \hline
98 & 58.5 & 8.6 & 40.6 & 7.9 & 26.6 & 4.0 \\ \hline
107 & 58.6 & 8.7 & 45.9 & 12.4 & 31.5 & 5.4 \\ \hline
118 & 67.7 & 8.7 & 49.6 & 19.5 & 36.9 & 7.4 \\ \hline
128 & 67.9 & 8.9 & 50.2 & 26.8 & 41.2 & 9.3 \\ \hline
140 & 72.8 & 8.8 & 48.3 & 36.0 & 45.6 & 11.9 \\ \hline
\end{tabular}
\end{table}
\indent In the present work, we have measured the K-shell x-ray yields of target atoms in three projectile-target systems, i.e., Si + Cu, Si + Zn, and Si + Ge. Using K x-ray yields, we have determined the K-shell ionization cross-section of the target atoms. It is observed that the present measurements are about a factor of two higher than the theoretical direct ionization cross-sections. We have employed the Fermi-Gas model (FGM) \cite{brandt1973dynamic} to determine the projectile charge-states inside the target material to explain the current findings in the light of electron capture induced target ionization \cite{lapicki1980electron}. We have found that  {the mean charge states predicted by FGM are close} to theoretical estimates  {by} the ETACHA4 code \cite{ETACHA4} provided the electron capture contribution is excluded. Using such mean charge states, the theoretical estimates of total ionization cross-section of targets are found to be in good agreement with the experimental measurements. Noteworthy here that the ETACHA4 code has never been used or suggested to obtain the charge state distribution inside the solid target. We explored the fact that it has got required potentiality to do so. We have validated our theoretical method in a different system, Si ion on Ti target at 0.3 to 0.7 MeV/u energies \cite{msimanga2016k}. Therefore, present work is appropriate enough for many applications as stated in the beginning. 
\begin{figure*}
\centering
\includegraphics[scale = 0.62]{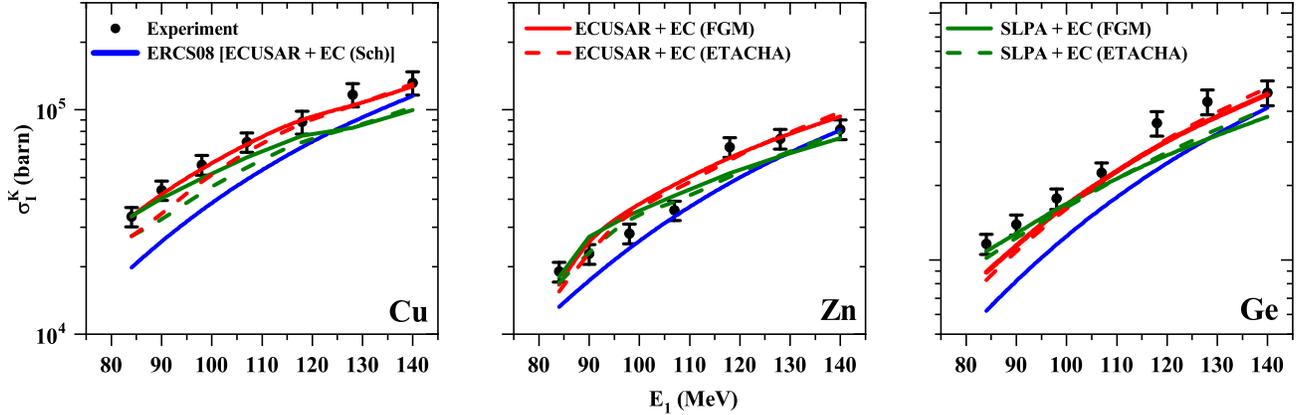}
\caption{Comparison of experimental K shell ionization cross-sections for different targets bombarded by the $^{28}$Si ions as function of ion-beam energies with the sum of the direct ionization and K-K + K-L capture cross-sections. Here, the direct ionization cross-sections are taken from ECUSAR \cite{lapicki2004effects}, SLPA \cite{montanari2013theory}, and ERCS08 \cite{horvat2009ercs08}. While the electron capture cross-section is calculated from Ref. \cite{lapicki1977electron} and the charge state fractions of the projectile ions inside the target are taken from (i) ETACHA4 \cite{ETACHA4}, (ii) Fermi gas model \cite{brandt1973dynamic}, which are denoted as EC(ETACHA4) and EC(FGM), respectively. Accordingly, we have represented different combinations of direct ionization and electron capture cross-sections in the figure.}
\label{Expt and theories} 
\end{figure*}
\begin{table*}[]
\centering
\caption{\label{CROSS-SECTION COMPARISON} Theoretical direct  ionization cross-sections obtained from ECUSAR ($DI_1$), SLPA ($DI_2$), and ERCS08 ($DI_3$) are given. Theoretical K-K capture cross-sections have been calculated using Ref.  \cite{lapicki1977electron} and for the Si projectiles of charge state 13+ and 14+.  Total capture cross-section taking charge state fraction (F(q)) of 13+ and 14+ from FGM \cite{brandt1973dynamic} (EC$_1$), ETACHA4 \cite{ETACHA4}  (EC$_2$) and ERCS08 \cite{horvat2009ercs08}  (EC$_3$) are also listed. Finally, sum of direct  ionization cross-section and total capture cross-section and experimental ionization cross-sections ($\sigma_{K}^I$) for ${Cu}$, ${Zn}$ and ${Ge}$ targets with about 10\% uncertainty  by the $^{28}{Si}$ ions of different energies ($E_1$ MeV) have been provided too. All the cross-sections are given in units of barns/atom.}
\begin{tabular}{|c|c|c|c|c|c|c|c|c|c|c|c|c|c|c|}
\hline
\multirow{2}{*}{\begin{tabular}[c]{@{}l@{}}\hspace{0.3cm}$E_1$\\ (MeV)\end{tabular}} & \multirow{2}{*}{$DI_1$} & \multirow{2}{*}{$DI_2$} & \multirow{2}{*}{$DI_3$} & \multicolumn{2}{l|}{\hspace{0.5cm}$EC_1$} & \multicolumn{2}{l|}{\hspace{0.5cm}$EC_2$} & \multirow{2}{*}{$EC_3$} & \multirow{2}{*}{\begin{tabular}[c]{@{}l@{}}$DI_1$ +\\ $EC_1$\end{tabular}} & \multirow{2}{*}{\begin{tabular}[c]{@{}l@{}}$DI_1$ +\\ $EC_2$\end{tabular}} & \multirow{2}{*}{\begin{tabular}[c]{@{}l@{}}$DI_2$ +\\ $EC_1$\end{tabular}} & \multirow{2}{*}{\begin{tabular}[c]{@{}l@{}}$DI_2$ +\\ $EC_2$\end{tabular}} & \multirow{2}{*}{\begin{tabular}[c]{@{}l@{}}$DI_3$ +\\ $EC_3$\end{tabular}} & 
\multirow{2}{*}{\begin{tabular}[c]{@{}l@{}}\hspace{0.3cm}$\sigma_K^I$\\ (Exptl.)\end{tabular}}\\ \cline{5-8}
& & &  & K-K  & K-L & K-K & K-L & & & & & & & \\ \hline
\multicolumn{15}{|l|}{\hspace{7.4cm}\textbf{Cu}} \\ \hline
\hspace{0.25cm}84                                                                     & 17080                   & 17280                   & 16400                   & 12241        & 4037         & 5974         & 4197         & 2728                    & 33358                                                                      & 27251                                                                      & 33558                                                                      & 27451                                                                      & 19808                                                                      & 33400                                                        \\ \hline
\hspace{0.25cm}90                                                                     & 21630                   & 19710                   & 20860                   & 15621        & 4997         & 7357         & 5179         & 4377                    & 42248                                                                      & 34166                                                                      & 40328                                                                      & 32246                                                                      & 26007                                                                      & 43800                                                        \\ \hline
\hspace{0.25cm}98                                                                     & 28400                   & 23110                   & 27460                   & 19945        & 6465         & 13438        & 6672         & 7575                    & 54810                                                                      & 48510                                                                      & 49520                                                                      & 43220                                                                      & 35975                                                                      & 56800                                                        \\ \hline
\hspace{0.25cm}107                                                                    & 36503                   & 27140                   & 35660                   & 25896        & 8283         & 20046        & 8593         & 12700                   & 70682                                                                      & 65141                                                                      & 61319                                                                      & 55778                                                                      & 49203                                                                      & 71800                                                        \\ \hline
\hspace{0.25cm}118                                                                    & 47540                   & 32340                   & 46500                   & 33385        & 10954         & 29700        & 11315         & 21250                   & 91879                                                                      & 88555                                                                      & 76689                                                                      & 73355                                                                      & 68790                                                                      & 88200                                                       \\ \hline
\hspace{0.25cm}128                                                                    & 58041                   & 37320                   & 56820                   & 32132        & 13692         & 31687        & 14072         & 30930                   & 103865                                                                     & 103800                                                                      & 83144                                                                      & 83079                                                                      & 88971                                                                      & 117000                                                       \\ \hline
\hspace{0.25cm}140                                                                    & 70864                   & 43550                   & 59460                   & 38751        & 17393        & 41658        & 17748        & 44300                   & 127008                                                                     & 130270                                                                     & 99694                                                                      & 102956                                                                      & 115164                                                                     & 132000                                                       \\ \hline
\multicolumn{15}{|l|}{\hspace{7.4cm}\textbf{Zn}}                                                                                                                                                                                                                                                                                                                                                                                                                                                                                                                                                                                                                                                                                   \\ \hline
\hspace{0.25cm}84                                                                     & 11617                   & 13030                   & 11330                   & 2042         & 2927         & 857          & 3030         & 1612                    & 16586                                                                      & 15504                                                                      & 17999                                                                      & 16917                                                                      & 13229                                                                      & 19000                                                        \\ \hline
\hspace{0.25cm}90                                                                     & 14825                   & 14900                   & 14500                   & 8654         & 3612         & 4607         & 3751         & 2612                    & 27091                                                                      & 23183                                                                      & 27166                                                                      & 23258                                                                      & 17437                                                                      & 22800                                                        \\ \hline
\hspace{0.25cm}98                                                                     & 19680                   & 17530                   & 19310                   & 11750        & 4696         & 11460        & 4851         & 4591                    & 36126                                                                      & 35991                                                                      & 33976                                                                      & 33841                                                                      & 24271                                                                      & 28100                                                        \\ \hline
\hspace{0.25cm}107                                                                    & 25819                   & 20640                   & 25400                   & 14665        & 6093         & 11321        & 6275         & 7842                    & 46577                                                                      & 43415                                                                      & 41398                                                                      & 38236                                                                      & 33661                                                                      & 35600                                                        \\ \hline
\hspace{0.25cm}118                                                                    & 34077                   & 24670                   & 33630                   & 19409        & 8041         & 17709        & 8298         & 13420                   & 61527                                                                      & 60084                                                                      & 52120                                                                      & 50677                                                                      & 47497                                                                      & 68200                                                        \\ \hline
\hspace{0.25cm}128                                                                    & 42082                   & 28540                   & 41610                   & 23602        & 10076         & 24088        & 10377         & 19920                   & 75760                                                                      & 76547                                                                      & 62218                                                                      & 63005                                                                      & 62002                                                                      & 74300                                                        \\ \hline
\hspace{0.25cm}140                                                                    & 52035                   & 33390                   & 51560                   & 28761        & 12803        & 32721        & 13145         & 29180                   & 93599                                                                      & 97901                                                                    & 74954                                                                      & 79256                                                                      & 81215                                                                      & 81800                                                        \\ \hline
\multicolumn{15}{|l|}{\hspace{7.4cm}\textbf{Ge}}                                                                                                                                                                                                                                                                                                                                                                                                                                                                                                                                                                                                                                                                                  \\ \hline
\hspace{0.25cm}84                                                                     & 5636                    & 7550                    & 5499                    & 1834         & 1433          & 1027         & 1603          & 583                     & 8903                                                                       & 8266                                                                       & 10817                                                                      & 10180                                                                       & 6219                                                                       & 11600                                                        \\ \hline
\hspace{0.25cm}90                                                                     & 7301                    & 8697                    & 7140                    & 2313         & 1787         & 1588         & 1933          & 957.2                   & 11401                                                                      & 10882                                                                       & 12797                                                                      & 12278                                                                      & 8258                                                                       & 13900                                                        \\ \hline
\hspace{0.25cm}98                                                                     & 9889                    & 10270                   & 9707                    & 3529         & 2368         & 2607         & 2599         & 1721                    & 15786                                                                      & 15095                                                                      & 16167                                                                      & 15476                                                                      & 11610                                                                      & 17700                                                        \\ \hline
\hspace{0.25cm}107                                                                    & 13269                   & 12170                   & 13060                   & 4588         & 3097         & 4194         & 3388         & 3028                    & 20954                                                                      & 20851                                                                      & 19855                                                                      & 19752                                                                      & 16297                                                                      & 22500                                                        \\ \hline
\hspace{0.25cm}118                                                                    & 17980                   & 14630                   & 17760                   & 6796         & 4207         & 6810         & 4523         & 5380                    & 28983                                                                      & 29313                                                                      & 25633                                                                      & 25963                                                                      & 23360                                                                      & 35800                                                        \\ \hline
\hspace{0.25cm}128                                                                    & 22709                   & 17010                   & 22480                   & 8525         & 5328         & 9779         & 5699         & 8267                    & 36562                                                                      & 38187                                                                      & 30863                                                                      & 32488                                                                      & 30976                                                                      & 43700                                                        \\ \hline
\hspace{0.25cm}140                                                                    & 28791                   & 20000                   & 28560                   & 11120        & 6896         & 13791        & 7305         & 12610                   & 46807                                                                      & 49887                                                                      & 38016                                                                      & 41096                                                                      & 41401                                                                      & 47600                                                        \\ \hline
\end{tabular}
\end{table*}
%
\indent 
\begin{figure*}[!t]
\centering
\includegraphics[scale = 0.62]{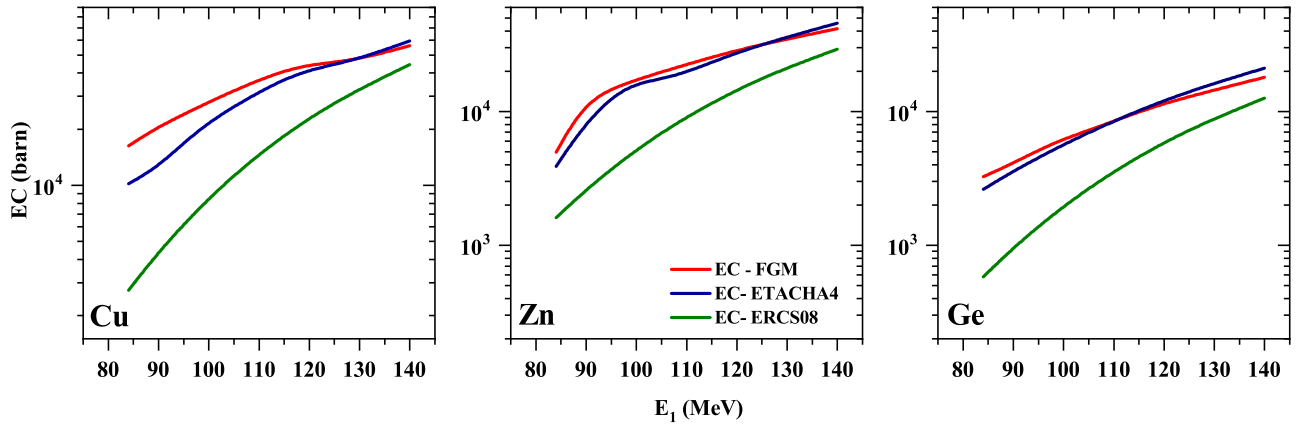}
\caption{Comparison of theoretical K-K capture cross-sections as calculated from Ref. \cite{lapicki1977electron} and the charge state fractions of the projectile ions are taken from (i) ETACHA4 \cite{ETACHA4},  (ii) Fermi gas model \cite{brandt1973dynamic} and (iii) ERCS08 \cite{horvat2009ercs08}. We have represented such combinations as EC-ETACHA4, EC-FGM and  EC-ERCS08, respectively.}
\label{EC comp}
\end{figure*}
\begin{figure*}[!h]
\centering
\includegraphics[scale = 0.43]{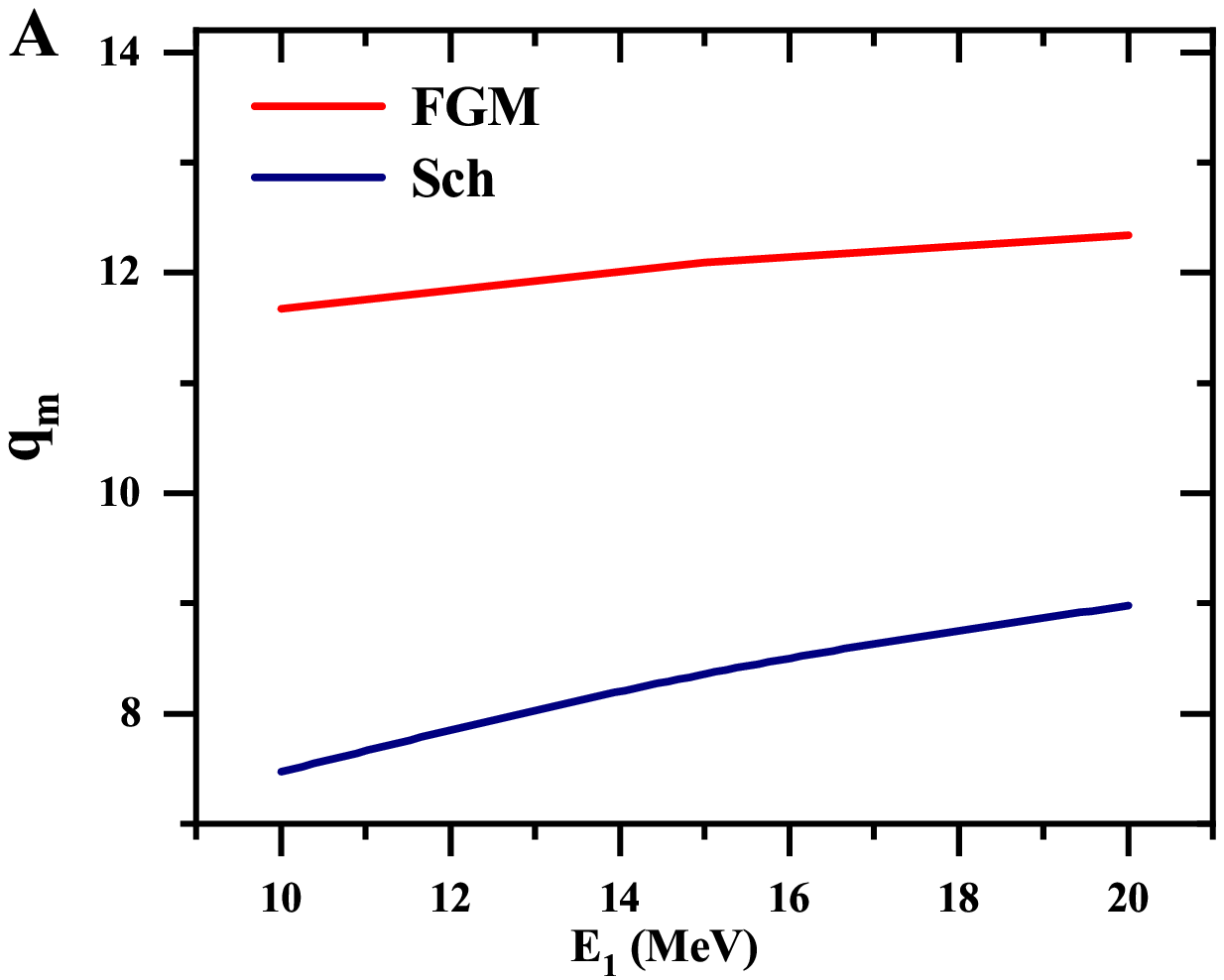}\includegraphics[scale = 0.43]{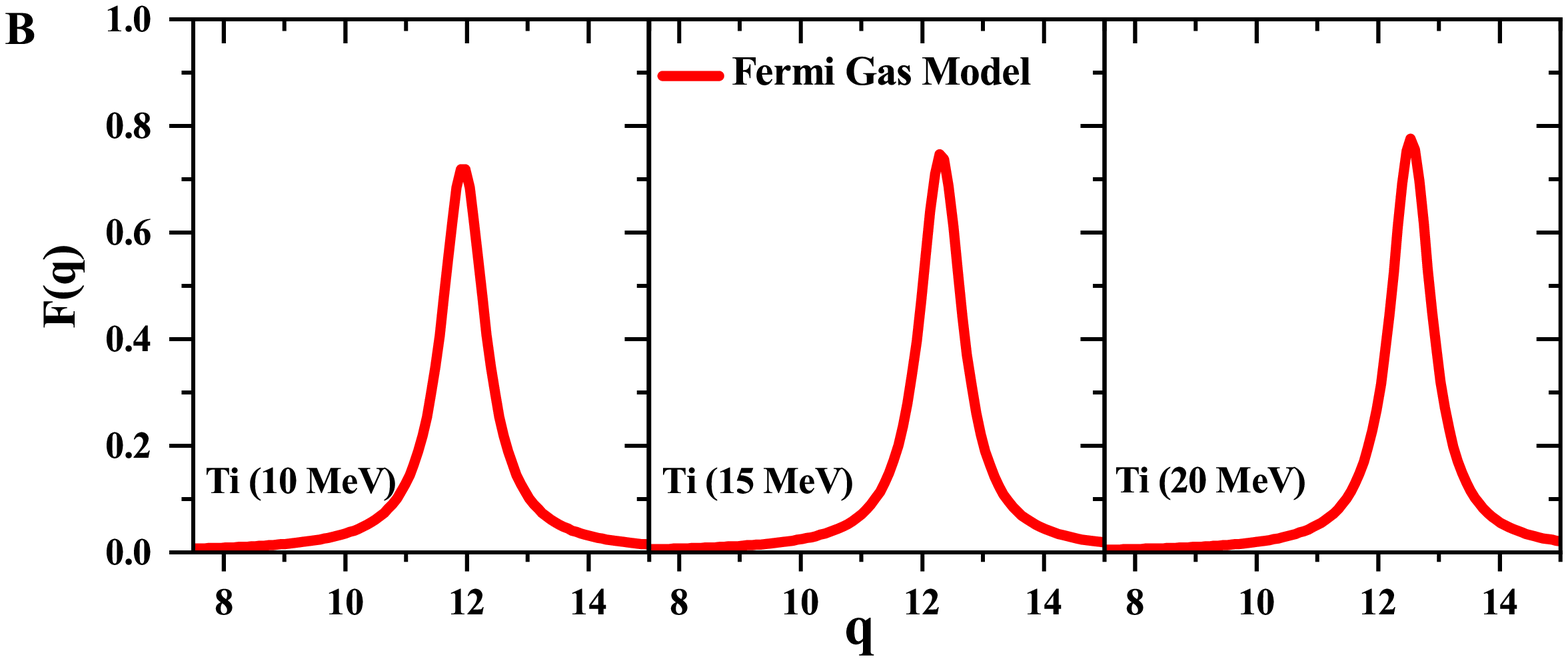}
\includegraphics[scale = 0.44]{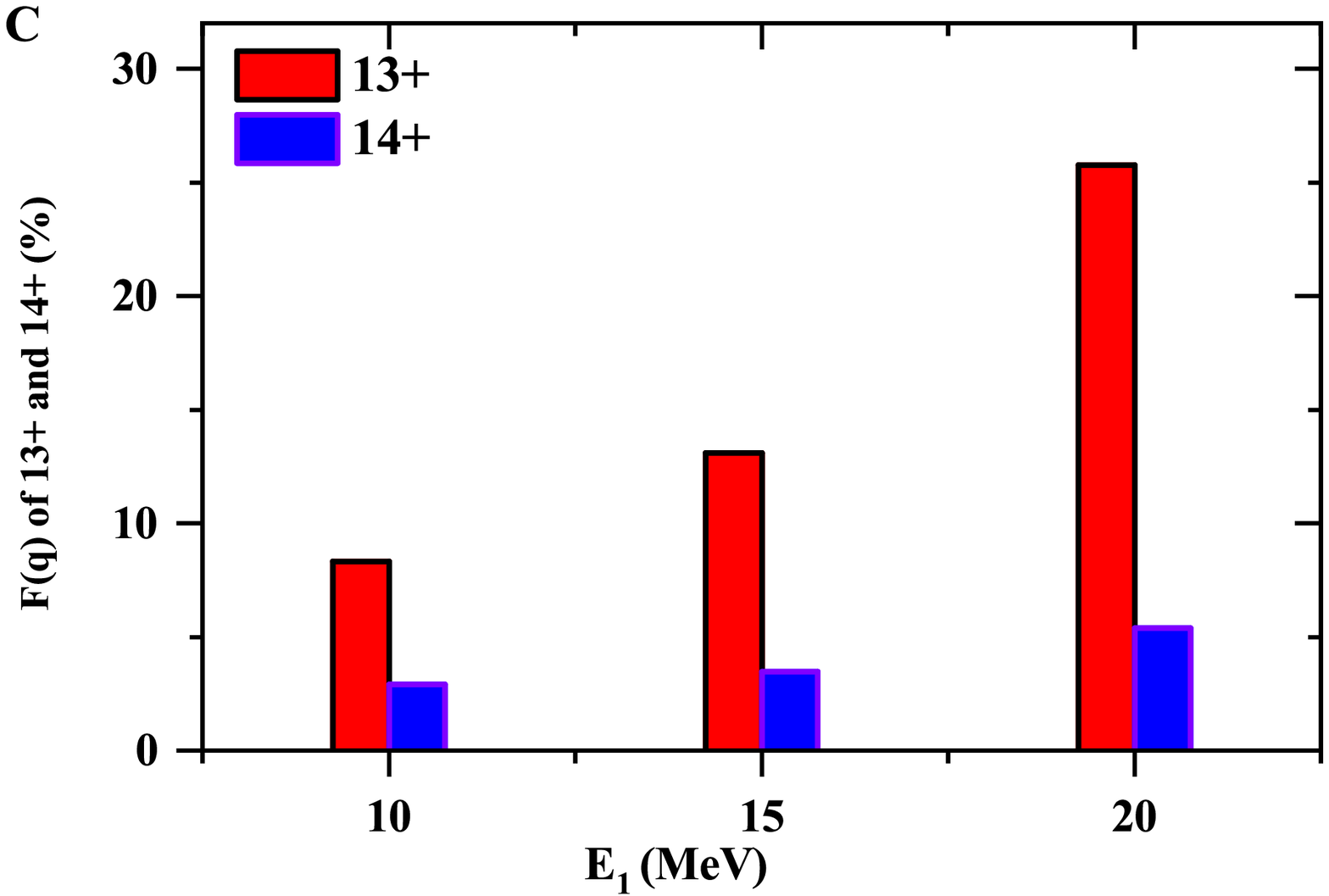}\includegraphics[scale = 0.44]{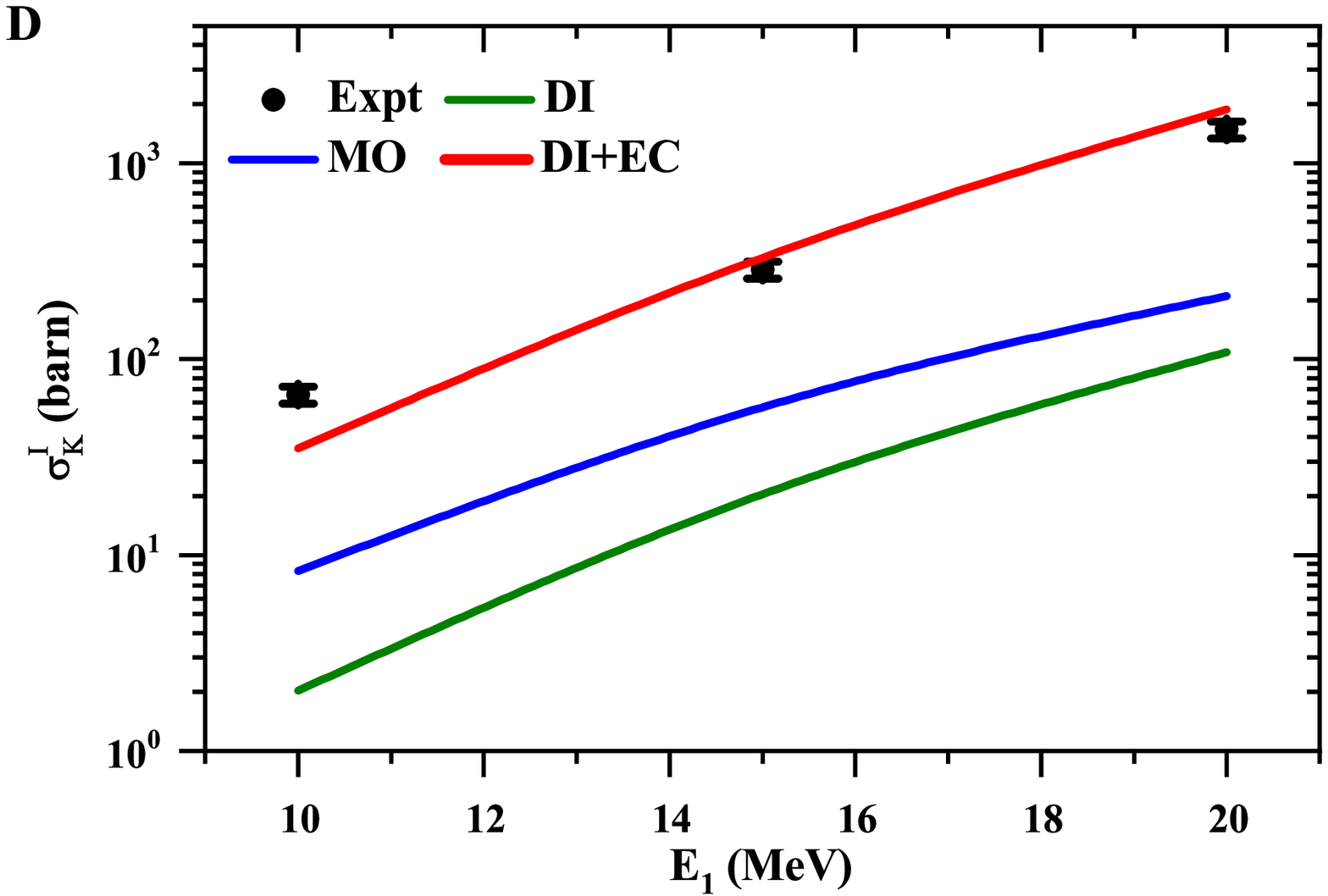}
\caption{(A)-(C) are same as Fig. \ref{Qm} and \ref{CSD BAR} but for ${Ti}$ target and data taken from \cite{msimanga2016k}. (D) comparison of experimental K shell ionization cross-sections with the ECUSAR (DI) \cite{lapicki2004effects}, molecular orbital (MO) cross-sections \cite{montenegro19851s} and the sum of the DI and electron capture (EC) cross-sections \cite{lapicki1977electron} versus ion-beam energies.}
\label{Si on Ti}
\end{figure*}

\section{Experimental details}
\indent The experiment was performed in the atomic physics beam line of 15 UD Pelletron which is situated at Inter-University Accelerator Centre, New Delhi (India). Si ion beam of charge state 8$^+$ for beam energies 84, 90, 98, 107 MeV) and charge state 12$^+$ for beam energies 118, 128, 140 MeV was obtained from Pelletron to bombard the natural Cu, Zn, and Ge targets. The vacuum of the order of $10^{-6}$ Torr was maintained in the chamber using turbo-molecular pump. Two silicon surface barrier detectors were placed at $\pm7.5^\circ$ with respect to beam direction to normalize the charge. A Si(Li) solid state detector was placed outside the chamber at $125^\circ$ with respect to beam direction and distance of 170 mm from the target. A collimator of 5 mm diameter was placed in front of the detector inside the chamber. The thickness of the Mylar window of the chamber for the detector was $6\mu m$. The specification of the detector (ORTEC, Oak Ridge, Tennessee, USA) is as follows: thickness 5 mm, diameter 10 mm, the thickness of Be window 25 $\mu m$ and energy resolution ~200 eV for Mn $K\alpha$ x-rays. The energy calibration of the detector was done before and after the experiment using the $^{55}$Fe, $^{57}$Co and $^{241}$Am radioactive sources.   {The target surface was placed at $90^{\circ}$ to the beam direction (normal to the target surface is collinear to the beam direction)} on a rectangular steel ladder which could move horizontal and vertical direction with the help of stepper motor. The spectroscopically pure (99.999\%) thin targets of natural Cu, Zn, and Ge were made on the carbon backing using vacuum deposition technique. The thickness of Cu, Zn, Ge, and carbon backing was 25 $\mu g/cm^2$, 14.4 $\mu g/cm^2$, 99 $\mu g/cm^2$, and 20 $\mu g/cm^2$, respectively\cite{singh2021kshell}. The thicknesses of targets were measured using the energy loss method using $^{241}$Am radioactive source. The data was acquired using a PC based software developed at IUAC \cite{subramaniam2010data}. The beam current was kept below 1 nA to avoid pile-up effects and damage to the targets. A semi-empirical fitted relative efficiency curve used for the present measurement can be seen in Oswal et.al. \cite{oswal2020experimental}.\\
\section{DATA ANALYSIS, RESULT, AND DISCUSSION}
\indent Typical K x-ray spectra of Cu, Zn, and Ge bombarded with 84 MeV and 140 MeV Si ions are shown in Fig. \ref{Spectra}. The spectra were analyzed with a nonlinear least-squares fitting method considering a Gaussian line shape for the x-ray peaks and a linear background fitting. The x-ray production cross-sections for the K x-ray lines were determined from the relation,
\begin{equation}
\sigma_i^x=\frac{Y_i^x A}{N_A\epsilon n_p t \beta} \label{eqn:3}
\end{equation}
\noindent here $Y_i^x$ is the intensity of the $i$th x-ray peak ($i$ = K$_\alpha$, K$_\beta$). $A$ is the atomic weight of the target. $N_A$ and $n_p$ denote the Avogadro number and the number of incident projectiles, respectively. $\epsilon$, $t$, and $\beta$ represent the effective efficiency of the x-ray detector, the target thickness in $\mu g/cm^2$, and the correction factor for energy-loss of the incident projectile and absorption of emitted x-rays in the target element, respectively.
The sum of $\sigma_{K_\alpha}^x$ and $\sigma_{K_\beta}^x$ gives a measure of the total K x-ray production cross-section as given in Table \ref{EXPERIMENTAL CROSS-SECTION}.\\

It is now well known that heavy ions produce simultaneous multiple ionization (SMI) in several shells while traveling through the target. SMI of L-shells along with a vacancy in K-shell will influence the value of K-shell fluorescence yield ($\omega_k$) to a considerable extent. Instead of using rigorous Hartree-Fock-Slater calculation for the $K_\alpha$ and $K_\beta$ peak shift due to the SMI effect, we employ a simple model of \textcite{burch1974simple}. According to it, the energy shift of $K_\alpha$ and $K_\beta$ lines per $2p$ vacancy with respect to corresponding diagram lines are 1.66$Z_L$ and 4.18$Z_L$ eV, respectively, where $Z_L$ = $Z_2$ -- 4.15; $Z_2$ is the atomic number of the target element. It is clear from Fig. \ref{Spectra} (a) that $K_\alpha$ and $K_\beta$ lines are well resolved for all the targets used in present measurements. In order to visualize the centroid shift due to the SMI effect, we have plotted the $K_\alpha$ and $K_\beta$ energies versus beam energy for all the targets in Fig. \ref{Spectra}(b). We notice, for all the targets, the corresponding centroid energies do not vary much with the beam energies used. The average K$_\alpha$ peak energies of $Cu$, $Zn$, and $Ge$ are 8.01 $\pm$ 0.1, 8.55 $\pm$ 0.1, and 9.87 $\pm$ 0.1 keV, respectively and are close to the corresponding diagram K$_\alpha$ lines at 8.03, 8.62, 9.86 keV.   {In contrast, this picture for K$_\beta$ lines is rather distinctive.} Mean of the measured K$_\beta$ lines 9.04 $\pm$ 0.13, 9.72 $\pm$ 0.12, and 11.2 $\pm$ 0.1 keV for Cu, Zn and Ge, respectively are higher than the corresponding diagram K$_\beta$ lines at 8.905, 9.572, and 10.982 keV. Thus, the difference between the measured $K_\beta$ and the diagram $K_\beta$ lines for Cu, Zn, and Ge are 135, 148, and 218 eV, respectively. These values are somewhat larger than the energy shift per $2p$ vacancy for $K_\beta$ lines, which are 104, 108, 116 eV, respectively, for Cu, Zn, and Ge. This figure along with the measurement uncertainty mentioned above implies that on the average two vacancies occur in $2p$ shells during the present collisions. This picture corroborates well the scenario in the $K_\alpha$ case too as the energy shift per $2p$ vacancy in Cu, Zn and $Ge$ for $K_\alpha$ line are only 41, 43, 46 eV, respectively, and the energy shift due to two $2p$ vacancies will be smeared in its measurement uncertainty of about 100 eV. Thus the SMI must be included in data analysis.\\
%
%
\indent Theoretically, K x-ray production cross-section ($\sigma_K^x$) can be obtained using the relation \cite{kadhane2003k}
\begin{equation}
\sigma^x_{K}= \omega_K\sigma^I_{K} \label{eqn:Sig-KI}
\end{equation}
here $\sigma^I_{K}$ is K-shell ionization cross-section, $\omega_k$ is the K shell fluorescence yield in the presence of SMI effect in L-shell. Single vacancy fluorescence yield $\omega_k^0$ given by Krause \cite{krause1979atomic} has been used. Hence, to extract the K shell ionization cross-section from the measured x-ray production cross-section one needs the accurate knowledge of $\omega_K$. To estimate it amidst the SMI process discussed above, we are following the description of Lapicki $et. al.$ \cite{lapicki2004effects} using an assumption that each electron in a manifold of outer subshells is ionized with an identical probability $P$ and correct $\omega_K$ in presence of SMI process becomes
\begin{equation}
 \omega_k=\frac{\omega_k^0}{1-P(1-\omega_k^0)}.
\end{equation}
With 
\begin{equation}
 P= q_m^2(1-\frac{0.225}{v_1^2})\times\frac{1}{1.8v_1^2}
\end{equation}
\noindent here $v_1=6.351 [E_1/A_1]^{1/2}$ ($E_1$ and $A_1$ are projectile energy and mass in MeV and amu units, respectively) is the projectile velocity. $q_m$ is the mean charge state of the projectile ion inside the target.\\
  {\indent The uncertainty in $\omega_k$ can be estimated from the following expression
\begin{multline}
\frac{\Delta\omega_k}{\omega_k}=\frac{\Delta\omega_k^0}{\omega_k^0}+\frac{P}{[1-P(1-\omega_k^0)]}\\
\times[\frac{\Delta P}{p}\times (1-\omega_k^0)-\frac{\Delta w_k^0}{\omega_k^0}]
\end{multline}
where
\begin{equation}
\begin{aligned}
\frac{\Delta P}{P}=\frac{2 \Delta q_1}{q_1}.
\end{aligned}
\end{equation}
The projectile velocity $v_1$ can be defined very precisely and thus its uncertainty is nominal ($< 1\%$) and taken as just a constant here. For Cu, $\frac{\Delta \omega_k^0}{\omega_k^0}$ is $\approx 5\%$ assume and this is $\approx 3\%$ for Zn and Ge. 
If we assume $\frac{\Delta q_1}{q_1} is \approx 3\%$ (its estimation and probable uncertainty will be discussed later), $\frac{\Delta \omega_k}{\omega_k}$ turns out to be $\approx 6\%$. }\\
\indent The inner-shell vacancies are produced predominantly by the direct Coulomb ionization process, which can be treated perturbatively using the first-order perturbation approaches, namely, the plane-wave Born approximation.
PWBA \cite{choi1973tables}. The standard PWBA approach for direct ionization were further developed to include the hyperbolic trajectory of the projectile, the relativistic wave functions, and the corrections for the binding-polarization effect. The most advanced approach based on the PWBA, which goes beyond the first-order treatment to include the corrections for
the binding-polarization effects within the perturbed stationary states (PSS) approximation, the projectile energy loss
(E), and Coulomb deflection (C) effects as well as the relativistic (R) description of inner-shell electrons, is known as
the ECPSSR theory \cite{brandt1981energy}. This theory is further modified to replace the PSS effect by a united and separated atom (USA) treatment and valid in the complementary collision regimes of slow and intermediate to fast
collisions, respectively \cite{lapicki2004effects}.\\
\indent The shell wise local plasma approximation (SLPA) \cite{montanari2011collective,montanari2013theory} is an ab-initio approach for the calculation of ionization probabilities within the dielectric formalism. It is based on the quantum dielectric response theory, generally employed to deal with the conduction band of solids, extended to account for the inner-shells by considering the density of target electrons and the binding energies. The SLPA calculates the  {K-shell} ionization cross-section of certain target atom due to the interaction with  {a projectile (velocity $v_1$ and nuclear charge $Z_1$)} as
\begin{equation}
\begin{aligned}
\sigma_{K}^{SLPA} = 2/(\pi v_1^2)\int_{0}^{\infty}\frac{Z_1^2}{p}dp\int_{0}^{p v_1}  d\omega\\
\int_{}^{}Im[\frac{-1}{\epsilon(p, \omega, E_K, \delta_K(r)}]\ \vec{dr},  
\end{aligned}
\end{equation}
 {with} $\epsilon$ $(p,\omega, E_K,\delta_K)$  {being the} Levine-Louie dielectric function \cite{levine1982new} {, $E_K$  the binding energy,  $\delta_K$ (r)  the density of the K-shell electrons around the nucleus, and p (w) the momentum (energy) transferred. } For Cu, Zn and Ge,  {we obtained} $E_K$, and $\delta_K(r)$  from the Roothaan-Hartree-Fock   {wave functions of neutral atoms} by Clementi and Roetti \cite{clementi1974roothaan}. These are the only inputs for our calculations. The SLPA has been successfully employed to describe the different moments of the energy loss of ions in matter, like ionization cross-sections of the L-shell \cite{oswal2018x, oswal2020experimental}, K-shell ionization \cite{kadhane2003k, kadhane2003experimental}, or mean energy loss \cite{montanari2020stopping, montanari2017low}.\\
\indent About a decade ago, \citet{horvat2009ercs08} developed  {a FORTRAN code (ERCS08) for computing the atomic electron removal cross-sections (ERCS).} The calculations are based on the ECPSSR theory for direct ionization  {and subsequent} modifications,  {while} the non-radiative electron capture is accounted by following \citet{lapicki1977electron,lapicki1980electron}.   {The ERCS08} program allows for selective inclusion or exclusion of individual contributions to the cross-sections. Thus, one can evaluate the K-shell ionization cross-section originated from direct ionization and nonradiative electron capture separately. \\
\indent In Fig. \ref{Expt and DI theories}, the measured $\sigma^I_{K}$ are compared with the predictions of direct ionization cross-section from ECUSAR, SLPA, and ERCS08,  {calculated} as mentioned above.  {As it can be noted, the} ECUSAR and ERCS08  {values} are almost equal, as expected for the present experimental conditions, while SLPA predictions are much lower than the other two. Whatsoever, with a great surprise, we see that measured $\sigma^I_{K}$ are at least a factor 2 higher than all the predictions. Note that the overall experimental uncertainty in the present cross-section measurements is attributed to the uncertainties in the photopeak, absolute efficiency of the detector, charge collected in Faraday cup, and target thickness.\\
\indent  {The underestimation of the experimental data by} the theoretical predictions provides a clear indication that the direct ionization process is not enough to explain the K-shell ionization phenomenon in the present experimental conditions and another important mechanism must be in action. Such a possibility can arise from electron capture phenomenon. K-K capture can be feasible if the K-shell of the projectile is either fully or partially vacant. Similarly, K-L capture will take place when L-shell of the projectile ion is unoccupied.\\
\indent To calculate the K-K electron capture cross-sections, we have used the theory of \citet{lapicki1977electron} which is based on the Oppenheimer-Brinkman-Kramers (OBK) approximation \cite{may1964formation} with binding and Coulomb deflection corrections at low velocities. Neglecting  {changes in the K-shell binding energy of the projectile} with one versus two  vacancies, a statistical scaling is used to  {relate} the electron transfer cross-section of one  ($\sigma_{1K \rightarrow K}$)  {and two ($\sigma_{2K \rightarrow K}$)  K-shell vacancies  as} $\sigma_{1K \rightarrow K} \sim \sigma_{2K \rightarrow  K}/2$. In the present experimental condition, $v_1= 10.96 - 14.15$ and $v_{2K}= 28.7 - 31.7$ atomic units.   {Thus,} the expresssion for $\sigma_{2K \rightarrow  K}$ can be  {chosen} as follows \cite{lapicki1977electron}:
\begin{equation}
\sigma_{2K \rightarrow K}=\frac{1}{3}\sigma_{2K\rightarrow K}^{OBK}(\theta_K), 
\label{KKCap} 
\end{equation}
\noindent  {where}
\begin{equation}
\sigma_{2K\rightarrow K}^{OBK}(\theta_K)= \frac{2^9 \pi a_0^2}{5v_1^2} \frac{(v_{1K}v_{2K})^5}{[v_{1K}^2+(v_1^2+v_{2K}^2-v_{1K}^2)^2/4v_1^2]^5},  
\label{KK_OBK}
\end{equation}
\begin{equation}
\theta_K=\frac{E_K}{Z_{2K}^2\times13.6},
\end{equation}
\noindent  {and} $Z_{2K}=Z_2-0.3$.
 {In Eq.(\ref{KK_OBK}),}  $a_0$, $v_{1K}$, and $v_{2K}$ are Bohr radius, K-shell orbital velocity for the projectile ion and target atom, respectively {; and}  $E_K$ is the binding energy of the K-shell electron of the target in eV. Note that this K-K capture theory has been adapted to estimate the K-L capture cross-sections too.  \\
\indent  {In this work we deal with $Si^{+14}$ and $Si^{+13}$, so the $\sigma_{1K \rightarrow K}$ and $\sigma_{2K \rightarrow K}$ are weighted with the charge state fractions F(q), for q=13 and q=14.}
To obtain the $F(q)$ we have used the following methods: (i) \textit{ab initio} approach by means of ETACHA4 code \cite{ETACHA4} and (ii) Fermi gas model based empirical formula \cite{brandt1973dynamic}.  {It is worth noting} that about a decade ago, the significance of the projectile charge state \textit{inside} the target on the target ionization was not known at all.   {Thus, \citet{horvat2009ercs08} has made use of the projectile charge state \textit{outside} the target, which is incorrect as it will be evident after a while.} \\
\indent  {The ETACHA4 code, recently developed by Lamour et al [53], computes} the charge state fractions of the projectile ions on the passage of a target medium, either solid or gas, by employing suitable rate equations. In the code, the non-radiative and radiative electron capture cross-sections are calculated using the relativistic eikonal approximation \cite{Meyerhof} and Bethe-Salpeter formula \cite{bethe1957quantum}, respectively. The total electron capture cross-section is sum of the non-radiative and radiative electron capture cross-sections. Whereas, the ionization and excitation cross-sections are estimated using the continuum distorted-wave-eikonal initial state approximation \cite{fainstein1987z,fainstein1991two} and symmetric eikonal model \cite{olivera1993electronic,ramirez1995distortion}, respectively. \\
\indent An important fact is that the excited states forming inside the solid target are mostly destroyed in the following collisions, in particular, if geometrical size  of the excited states so created is larger than the lattice parameter of the target material. Whereas significant contribution of electron capture at the exit layers remains intact.   This is the reason,  the excited state formation is considered to be occurring at the exit surface \cite{tolk1981role}. Hence, putting the electron capture cross-section equals to zero in the  {ETACHA4 code, provides}   {us a good estimate of the CSD inside the solid target (CSD-I). This is important}  {as CSD-I will be} {used later on to calculate the electron capture contribution in the K-shell ionization in the target atoms.}\\ 
\indent  {Instead,} \citet{horvat2009ercs08} in his ERCS08 code used an empirical formulae for the mean charge state, {$q_m$,}  as well as CSD outside the target, {CSD-O, by}  \citet{schiwietz2001improved}. We  {emphasize} here that this empirical  {values outside the target} do not represent the quantities inside the target at all.\\   
\indent According to the  Fermi gas model based empirical formula, the mean charge state ($q_m$) inside the target \cite{brandt1973dynamic} is given by,
\begin{equation}
q_m = Z_1(1- {v_F\over v_1})
\end{equation}
 {with} $Z_1$ and $v_F$  {being} the projectile atomic number and Fermi velocity of target electrons, respectively.   {Series of $q_m$-values obtained from x-ray spectroscopy experiments have been compared extremely well with the the above-mentioned formula \cite{nandi2019unusual}. Uncertainty of $q_m$ is found to be $\approx 3\%$.} The Fermi velocity  $v_F$ of Cu, Zn, and Ge are $1.11\times10^6$, $1.566\times10^6$ \cite{gall2016electron} and $2.5\times10^6$ m/s \cite{isaacson1975compilation}, respectively.\\
\indent   {To showcase the difference of ionization of the projectile ion inside and outside the target, we compare the $q_m $ obtained from the Fermi gas model \cite{brandt1973dynamic} with the empirical model by \citet{schiwietz2001improved} (Schiwietz-Grande model). The Schiwietz-Grande model was developed from a large set of experimental charge-state distributions  {measured outside the solid target}. We displayed the $q_m $ as predicted by the Fermi gas model \cite{brandt1973dynamic}, and Schiwietz-Grande model \cite{schiwietz2001improved} in Fig. \ref{Qm}. This contrasting picture is mostly governed by the solid surface \cite{nandi2008formation,sharma2019disentangling}.}  {We also included in Fig. \ref{Qm} the results of ETACHA4 code inside and outside the target. Clearly, the values of $q_m$ outside the target are lower than inside in all the energy range studied here.} \\
\indent In second step, the $q_m$-values inside the target are substituted in the Lorentzian charge state distribution \cite{sharma2016experimental} to obtain the $F(q)$ as follows
\begin{equation}
F(q)=\frac{1}{\pi}\frac{\frac{\Gamma}{2}}{(q-q_m)^2+(\frac{\Gamma}{2})^2} \:\text{and}\:\sum_q F(q)=1. \label{CSF}
\end{equation}
\noindent Here distribution width $\Gamma$ is taken from Novikov and Teplova \cite{novikov2014method}, as follows,
\begin{equation}
\centering
\Gamma(x)= C[1-exp(-(x)^\alpha)][1-exp(-(1-x)^\beta)]\label{Gamma}.
\end{equation}
Here $x=q_m/Z_1$, $\alpha=0.23$, $\beta=0.32$, and $C=2.669-0.0098× Z_2+0.058×Z_1+0.00048×Z_1× Z_2$. The $F(q)$ values so obtained are shown in Fig. \ref{CSD BAR}(A) and the $F(q)$ for q=13 and 14 are displayed in a bar chart Fig. \ref{CSD BAR}(B). Further, charge state fraction (F(q)) of Si$^{13+}$ and Si$^{14+}$ obtained from FGM \cite{brandt1973dynamic}, ETACHA4 \cite{ETACHA4}, and ERCS08 \cite{horvat2009ercs08} in different target elements and at various kinetic energies of Si-ion beam are given in Table \ref{CSD FRACTION}. Note that FGM and ETACHA4 represent $F(q)$ inside the target, while ERCS08 gives the same outside the target, hence $F(q)$ from ERCS08 is not at all useful in understanding the vacancy production in target atoms by ion impact.\\
\indent The $\sigma_{KK}$  and $\sigma_{KL}$ so obtained were added with the $\sigma^I_{K}$ as obtained from direct ionization theories and plotted in Fig. \ref{Expt and theories}.  {The improvement of the theoretical-experimental comparison from Fig. \ref{Expt and DI theories} (without capture) to Fig. \ref{Expt and theories} (ionization plus K-K and K-L capture) is very clear.} These data are also given in Table \ref{CROSS-SECTION COMPARISON}. The sum of direct  {ionization} and K-K + K-L capture cross-sections show a good agreement with the corresponding experimental cross-sections.  {As can be noted in Fig. \ref{Expt and theories}, ETACHA and FGM for this addition almost agree for Zn and Ge target, with differences for Cu.} Furthermore, ECUSAR - FGM gives the closest agreement with the experimental data. Such  {overall} agreement reveals that a simple Fermi gas model gives a correct estimation of the $q_m$ inside the target, where $v_F$ plays a central role and needs accurate evaluation.\\ 
\indent   {Due to certain CSD inside the target, the charge state fraction for a specific $q$ called $F(q)$ is an important quantity. The effective 2K-K OBK capture contribution in the present case is then equal to $F(q = 14)\times \sigma_{2k\rightarrow k}^{OBK}(\theta_k)$, for silicon ions. Similarly, to obtain effective K-K contribution F(q = 13) will be required.} Note that ERCS08 \cite{horvat2009ercs08} code takes the charge state fractions of the projectile ions outside the target \cite{schiwietz2001improved}\\
\indent  {We tested the above mentioned approach by reanalysing earlier data for C- and Si-ion on Ti target at much lower energies. Details and the excellent results obtained are included in Appendix \ref{A}, and reinforce the present conclusions. } \\
\indent To  {deepen our study about} which theoretical method gives the best representation of the experimental data, we have compared the total electron capture cross-section obtained from different theoretical approaches in Fig. \ref{EC comp}. Here, we see that the electron capture cross-sections almost follow the mean charge state behavior shown in Fig. \ref{CSD BAR}. Further,  the electron capture cross-sections are close if the charge state fractions are taken from FGM \cite{brandt1973dynamic} and ETACHA4 \cite{ETACHA4}. Therefore, either FGM or ETACHA4 can be used to estimate the projectile charge state distribution inside the solid target. {However, ETACHA4 can handle up to a certain number of electrons in the projectile ion and thus difficulties arise in applying this for heavy projectiles. Whereas no such restrictions with the FGM.} Electron capture cross-sections obtained from the ERCS08 is very low as the charge state distribution outside the target is considered there.\\ 
\section{Conclusion}
\indent We have demonstrated the contribution of electron capture in K-shell ionization by heavy-ion impact. Here, the targets Cu, Zn, and Ge were bombarded by the 84-140 MeV $^{28}$Si ions to measure K-shell production cross-sections. We observed that the measured ionization cross-sections differ at least a factor of two from the theoretical direct ionization cross-sections. Electron capture from the target K-shell to the K- and L-shell of the projectile ions was required to resolve this difference. In this regard, projectile charge state inside the target is extremely essential. Use of mean charge state from a Fermi gas model \cite{brandt1973dynamic} and distribution width from Novikov and Teplova formula \cite{novikov2014method} in the Lorentzian charge state distribution \cite{sharma2016x} lead to obtaining the charge state distributions. The bare and H-like projectile ions inside the targets have been utilised to calculate the K-K capture contribution from \citet{lapicki1977electron}. While all the charge state fractions are used for K-L capture calculations. Sum of the theoretically calculated direct ionization cross-section and K-K + K-L capture cross-sections represent pretty well the experimentally measured values.\\
\indent {In order to validate the above mentioned approach, we have reanalysed the earlier data for C- and Si-ion on Ti target at much lower energies.} Hence, in this study we have not only succeeded in studying the dynamics of K-shell ionization, but also succeeded in correctly estimating the charge state distribution inside the targets using a simple Fermi gas model. Further, we have shown that the mean charge state inside the foil is much higher than that outside it. Thus, the exit surface plays an significant role in changing charge state from a higher to lower one.   {Such a knowledge can be put into application to obtain a high charge state from a foil stripper if the electron capture phenomena at the stripper surface is restricted. Quasi-free electrons at the conducting surface can be captured easily with the exiting ions. Whereas such scope is remote from the insulating surface. Hence, a special surface engineering can be applied to make a target surface from conducting to insulating one for obtaining the higher charge states.}
\section{acknowledgements}
\indent We acknowledge cooperation from the Pelletron accelerator staff during the experiments. SC acknowledges the University of Kalyani for providing generous funding towards his fellowship. CCM acknowledges the financial support by CONICET and PICT2017-2945 from Argentina. 
\appendix
\setcounter{secnumdepth}{0}
\section{Appendix}
\section{Validation of the theoretical method} \label{A}
The theoretical approach so developed has been validated through earlier experimental results of Si ion on Ti \cite{msimanga2016k} and shown in Fig.  \ref{Si on Ti} and Table \ref{SI_ON_TI}. It is observed there \cite{msimanga2016k} that C-data agree well with ECPSSR predictions \cite{brandt1979shell} but Si-data are 30 to 15 times higher than the ECPSSR estimations in the energy range of 10-20 MeV. Inclusion of SMI with ECUSAR calculation for accounting direct ionisation exhibits nearly the same scenario, but molecular orbital (MO) theory \cite{montenegro19851s} improves the results to a some extent at this low energies $<$ 1 MeV/u, but still away from the observed scenario. Next, to take K-K and K-L capture cross-sections \cite{lapicki1977electron} into consideration, we find the $q_m$ inside the target by the Fermi gas model \cite{brandt1973dynamic}. Values so obtained are compared with that outside the target using the Schiweitz model \cite{schiwietz2001improved} is shown in Fig. \ref{Si on Ti} A. Note that Fermi velocity of Ti is taken as 1.38$\times 10^6$ m/s \cite{isaacson1975compilation}. The CSD inside the target is depicted at different energies in Fig. \ref{Si on Ti} B. Corresponding charge state factions for $q=13+$ and $14+$ as shown in Fig. \ref{Si on Ti} C are used in the K-K capture cross-section \cite{lapicki1977electron}. Finally, different theoretical values are compared in Fig. \ref{Si on Ti} D, where the sum of the K-K + K-L capture cross-section \cite{lapicki1977electron} and the ECUSAR cross-section agrees within 25\%  with the experimental data except a larger difference ($\approx$ 45\%) at the lowest energy where the photopeak is very weak and thus background might have been underrated. Further, measurement uncertainty is not quoted there. Whatsoever, this excellent agreement is achieved because of 30-15 times higher contribution from electron capture than that from direct ionization in Ti atoms comes as we move from 10 to 20 MeV energy of Si-projectiles. 
\begin{table}[ht]
\centering
\caption{\label{SI_ON_TI} Theoretical direct ionization cross-section obtained from ECUSAR including simultaneous multiple ionization ($\sigma_{DI}$), molecular orbital (MO) cross-section ($\sigma_{MO}$) \cite{montenegro19851s}, K-K capture cross-section \cite{lapicki1977electron} for 13+ ($\sigma_{K-K}^{13+}$) and 14+ charge state ($\sigma_{K-K}^{14+}$) from FGM, total capture cross-section plus MO cross-section  ($\sigma_{Tot}=\sigma_{DI}$+$\sigma_{K-K}$) and experimental ionization cross-section ($\sigma_{K}^I$) for Ti target by the $^{28}$Si ions of different energies ($E_1$ MeV) \cite{msimanga2016k} are listed in units of barns/atom. Note that measurement uncertainty is not mentioned there.}
\begin{tabular}{|l|l|l|l|l|l|l|}
\hline
\multicolumn{7}{|l|}{\hspace{0.76cm}Ti (Experimental data from \cite{msimanga2016k})} \\\hline
\multirow{2}{*}{\begin{tabular}[c]{@{}l@{}}\hspace{0.1cm}$E_1$\\  MeV\end{tabular}} & \multirow{2}{*}{$\sigma_{DI}$} & \multirow{2}{*}{$\sigma_{MO}$} & \multicolumn{2}{l|}{\hspace{0.45cm}EC} & \multirow{2}{*}{$\sigma_{Tot}$} & \multirow{2}{*}{\begin{tabular}[c]{@{}l@{}}\hspace{0.57cm}$\sigma_K^I$\\     \hspace{0.25cm}(Exptl.)\end{tabular}} \\ \cline{4-5}
& & & K-K & K-L &  &  \\ \hline
10 & 2 & 8 & 16 & 17 & 35 & ~~~~~66 \\ \hline
15 & 20 & 57 & 235 & 117 & 372 & ~~~~286 \\ \hline
20 & 108 & 209 & 1412 & 354 & 1874 & ~~~1481\\ \hline
\end{tabular}
\end{table}
\bibliography{BIBLIO.bib}
\bibliographystyle{apsrev4-1}
\end{document}